\documentclass[12pt]{article}

\input{style.sty}

\newcommand{\pvarConstructIndividualsNAll}{502,507\ }
\newcommand{\pvarConstructIndividualsNNonmissingLocDate}{444,707\ }
\newcommand{\pvarAnalysisObsInSmallCohorts}{48\ }

\newcommand{\pvarAnalysisObsDroppingNoBF}{339,385\ }
\newcommand{\pvarAnalysisObsDroppingNoDistrict}{308,847\ }
\newcommand{\pvarAnalysisObsSiblingTwinsDropped}{250\ }
\newcommand{\pvarAnalysisObsSiblingSample}{23,500\ }
\newcommand{\pvarAnalysisObsFamsSiblingSample}{11,431\ }
\newcommand{\pvarAnalysisRsqBFbetween}{0.73\ }
\newcommand{\pvarAnalysisRsqBFwithin}{0.27\ }

\usepackage{afterpage}

\begin{document}

\title{The long-run returns to breastfeeding\thanks{We gratefully acknowledge financial support from the European Research Council (ERC StG 851725) and the European Union (ESSGN, project no. 101073237). This research has been conducted using the UK Biobank Resource under Application Number 74002. The views expressed in this publication are those of the authors and do not necessarily reflect those of the European Union, MSCA Horizon Europe, or ESSGN. Neither the European Union, nor the granting authority or ESSGN can be held responsible for them.}}

\date{\today \vspace{-2em}}

\author{Marco Francesconi\thanks{Department of Economics, University of Essex. E-mail: \href{mailto:mfranc@essex.ac.uk}{mfranc@essex.ac.uk}} \and Stephanie von Hinke\thanks{Corresponding author. School of Economics, University of Bristol; Institute for Fiscal Studies. E-mail: \href{mailto:S.vonHinke@bristol.ac.uk}{S.vonHinke@bristol.ac.uk}} \and Emil N. S\o{}rensen\thanks{School of Economics, University of Bristol. E-mail: \href{mailto:E.Sorensen@bristol.ac.uk}{E.Sorensen@bristol.ac.uk}}}

\maketitle

\begin{abstract}
\singlespacing\vspace{-1.5em}
\noindent 
This paper shows that the mid-20$^{th}$ century was characterised by a considerable reduction in breastfeeding rates, reducing from over 80\% in the late 1930s to just over 40\% only three decades  later. We investigate how maternal breastfeeding during this period has shaped offspring health and human capital outcomes in the UK. We use a within-family design, comparing children who were breastfed to their sibling(s) who were not. Our results show that breastfeeding increases adult height, as well as fluid intelligence, but does not affect educational attainment, nor adult BMI. In further analyses, we examine whether and how this impact varies with individuals' genetic ``predisposition'' for these outcomes, proxied by the outcome-specific polygenic index. We find that the ``height-returns'' to breastfeeding are larger among those genetically predisposed to be taller, with no genetic heterogeneity for the other outcomes, though we note that power in the within-family $G\times E$ analysis is more limited. Overall, our estimates suggest that breastfeeding plays an non-negligible role in child development.\\[3mm]
\end{abstract}

\clearpage

\section{Introduction}
In the early 1900s, around 85\% of babies in England were breastfed in their first two months of life, but this steadily declined during the second half of the 20$^{th}$ century, with the nadir in breastfeeding rates observed in the 1970s \citep{crowther2009resurgence}. 
Despite the substantial drop in rates of breastfeeding, in particular from the 1940s onwards, there has been little coverage of this period in the literature, which tends to focus on more recent times. This is likely to be partially driven by a general absence of consistently reported historical data on maternal breastfeeding. This paper addresses this gap, showing the prevalence, trends and heterogeneity in maternal breastfeeding rates in the UK, focusing on women whose children were born between the late 1930s and early 1970s. Between 2006 and 2010, these children were asked whether they were breastfed, allowing us to get an interesting insight into rates of breastfeeding during the war and post-war reconstruction period. This is the first contribution of our paper. 

The second contribution speaks to the longer-term impacts of breastfeeding on offspring development. Numerous studies highlight the positive association between maternal breastfeeding and offspring outcomes, with (longer durations of) breastfeeding generally associated with lower cholesterol and obesity levels, as well as reduced probabilities of diabetes, asthma and hypertension \citep[for reviews, see e.g.][]{hoddinott2008breast, horta2015long, victora2016breastfeeding, patro2019duration}, and improved cognitive outcomes \citep{walfisch2013breast, mcgowan2023benefits}. 
Although the literature acknowledges the potential for confounding in the relationship between breastfeeding and child outcomes, it is notoriously difficult to account for this. Indeed, breastfeeding is a choice, which has been shown to correlate with e.g., parental education, income and social class \citep[see e.g.,][]{heck2006socioeconomic, oakley2013factors, meek2022policy}. To deal with this, much of the literature uses covariate adjustment, controlling for potential confounders that tend to reduce estimated effect sizes \citep[see e.g.,][]{horta2015long}, or matching to account for the fact that mothers who breastfed may be systematically different from those who do not \citep[see e.g.,][]{belfield2012benefits, rothstein2013breastfeeding, borra2012effect}. 
A systematic review and meta-analysis of breastfeeding promotion interventions (including randomized but also non-randomized intervention studies) finds significant but modest improvements in child anthropometric measures \citep{giugliani2015effect}. The only randomized controlled trials that we are aware of (promoting breastfeeding in Belarussian hospitals) showed no evidence that increased durations of breastfeeding affected child adiposity, despite substantial increases in the duration and exclusivity of breastfeeding \citep{kramer2007effects, martin2013effects}.

More recent studies have used non-experimental causal inference methods to deal with confounding. For example, \citet{fitzsimons2022breastfeeding} exploit variation in breastfeeding driven by the \textit{timing} of births, showing that breastfeeding rates are lower among those giving birth just before or early into the weekend, arguably because of reduced infant feeding support. Similarly, \citet{del2012breastfeeding} leverage the roll-out of a breastfeeding support initiative in  UK hospitals, using the distance between the mother's home and the closest hospital as an instrument for the breastfeeding decision, and \citet{lawler2024effect} study the effect of hospital regulations that intended to increase breastfeeding rates. \citet{baker2008maternal} use an IV approach, exploiting variation in breastfeeding driven by increases in maternity leave mandates in Canada. This literature tends to find that breastfeeding improves child cognitive development, with no (or small) impacts on child health. Finally, there is a handful of studies that use within-family / sibling analysis. For example, \citet{rothstein2013breastfeeding} and \citet{colen2014breast} find large associations between breastfeeding and child outcomes, which generally become insignificant once family fixed effects are included, suggesting that the positive association is driven by family and environmental factors rather than breastfeeding \textit{per se}. 

We add to this literature in two main ways. First, we estimate the long-term health and human capital returns to breastfeeding both \textit{between} and \textit{within} families. For the latter, we compare the outcomes of children who were breastfed to those of their siblings who were not, focusing on anthropometrics (height, BMI) as well as cognitive skills (fluid intelligence, years of education) in older age. Our within-family design ensures that the family environment is held constant, exploiting only variation between siblings. Although such designs are unlikely to fully eliminate any confounding, it should substantially reduce this concern \citep{smithers2015effects, harden2021}. 

Second, we directly incorporate molecular genetic data into our empirical analysis, allowing us to explore potential genetic heterogeneity in the returns to breastfeeding. More specifically, we investigate whether the effects of breastfeeding differ for those differentially ``predisposed'' to the outcomes of interest, as captured by a polygenic index (PGI) that is predictive of the outcome, allowing us to explore whether one's genetic variation exacerbates or mitigates the returns to breastfeeding. This analysis speaks to the literature in medical as well as social science genetics that examines the role of the nature--nurture interplay in shaping individuals' outcomes \citep[see e.g.,][]{Rutter2006, biroli2025}. Our setting has two particular advantages. One, in addition to reducing confounding factors in breastfeeding, our within-family setting allows us to estimate direct (causal) genetic effects, and investigate whether and how these vary with maternal investments. Two, our setting speaks to economic theory that emphasizes the potential for complementarities between endowments and investments \citep{cunha2007technology}, as well as sociological theory, mirroring agency and structure models \citep{mills2020sociology}. We follow \citet{muslimova2025nature} and use \textit{within}-family genetic information as a proxy for such endowments. This allows us to explore whether the impact of breastfeeding on e.g. human capital is stronger for those who randomly inherited higher ``genetic endowments'' for human capital than their siblings. We consider these within-family $G\times E$ analyses as more suggestive however, due to a reduction in power driven by the within-family design.

\section{Results}
We run our analyses using the UK Biobank \citep{sudlow2015,bycroft2018uk}, a prospective population-based cohort of approximately 500,000 individuals aged 40 to 69 who resided in the UK at the time of recruitment, between 2006 and 2010. The data and our sample selection are discussed in more detail in \textit{Supplementary Information (SI)} Appendix \ref{sec:data}, as well as in, e.g., \citet{Fry2017}.

\begin{figure}[!h]\caption{\label{fig:bf_by_cohort_ukb}Breastfeeding by cohort}\centering\includegraphics[width=0.65\textwidth]{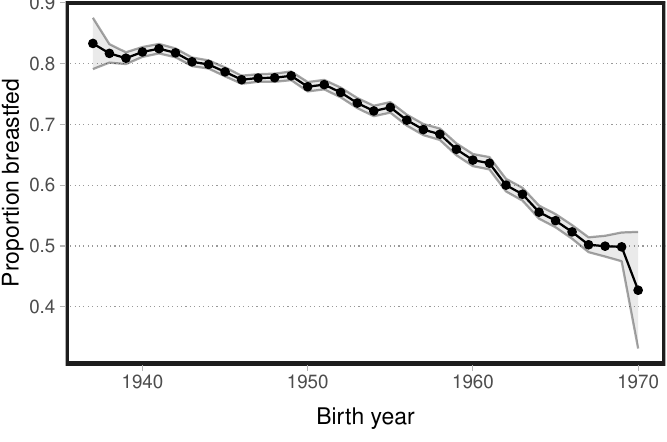}\caption*{\emph{Plots proportion (with 0.95 confidence interval) of UK Biobank participants that were breastfed by year of birth.}}\end{figure}

We start by plotting the rates of breastfeeding during this period. As shown in \autoref{fig:bf_by_cohort_ukb}, over 80\% of UK children born in the late 1930s reported to have been breastfed, plummeting during our observation period, to just $\sim$40\% of those born in the late 1960s. \textit{SI Appendix} \ref{sec:descr} describes the data in more detail. This also presents raw associations between breastfeeding and our outcomes of interest, showing that individuals who were breastfed as infants on average have higher fluid intelligence, more years of education, are taller, but with lower BMIs (\autoref{fig:v2_long-run-by-bf_raw}; \textit{SI}). The difference in human capital outcomes tends to increase for younger cohorts, whereas height differences show some convergence.

We further illustrate substantial regional variation in breastfeeding (\autoref{fig:bf-by-district}; \textit{SI}), as well as a strong social gradient, with breastfeeding being more common in higher SES areas (\autoref{fig:bf-by-ses}; \textit{SI}), defined as districts with an above median share of residents in professional, managerial or technical occupations, according to the 1951 UK Census. 
The latter highlights the importance of confounding in the relationship with offspring outcomes. Consistent with this, we show that individuals who were breastfed on average had higher birth weights compared to those who were not breastfed, and their mothers were less likely to smoke around the birth of their child. Similarly, they have higher PGIs for education and fluid intelligence. Most of these relationships attenuate and become insignificant however, once we only exploit variation between siblings within the same family (\autoref{tab:assoc_observables}, \textit{SI}). 

We next investigate the association between breastfeeding and individuals' later-life anthropometric and human capital outcomes. We distinguish between the full sample and the sibling sample, where the latter allows us to reduce confounding driven by e.g., family socio-economic characteristics. Appendix \ref{sec:power_analysis} \textit{(SI)} assesses the power of our analyses, showing that at 80\% power in the sibling sample, we can reliably detect effect sizes as small as 0.21 (BMI), 0.41 (height), 0.10 (educational attainment), and 0.09 (fluid intelligence). In the within-family design, the corresponding detectable effect sizes are 0.33 (BMI), 0.62 (height), 0.14 (educational attainment and fluid intelligence), which are larger because identification relies only on within-family variation. While the pooled specification is therefore more statistically powerful, the pooled and within-family specifications need not identify the same estimand, as the former may be confounded by shared family factors whereas the latter isolates within-family variation.

Panel A in \autoref{tab:v2_assoc_long-run-outcomes} presents the estimates from a linear regression of the outcome on a binary indicator for having been breastfed, controlling for gender, year-month of birth fixed effects, and local area (district) of birth fixed effects. This confirms the raw associations presented in \autoref{fig:v2_long-run-by-bf_raw} (\textit{SI}), showing that those who were breastfed have significantly higher intelligence and educational attainment. They are also taller, but have lower BMIs. 

\FloatBarrier
\begin{table}[!h]
\centering\centering\centering
\caption{\label{tab:v2_assoc_long-run-outcomes}Associations between long-run outcomes and breastfeeding.}
\centering
\begin{threeparttable}
\fontsize{10}{12}\selectfont
\setlength{\tabcolsep}{10pt}
\begin{tabular}[t]{ldddd}
\toprule
\multicolumn{1}{c}{\em{}} & \multicolumn{4}{c}{\em{Dependent variable:}} \\
\cmidrule(l{3pt}r{3pt}){2-5}
\multicolumn{1}{c}{} & \multicolumn{1}{c}{(1)} & \multicolumn{1}{c}{(2)} & \multicolumn{1}{c}{(3)} & \multicolumn{1}{c}{(4)} \\
\multicolumn{1}{c}{ } & \multicolumn{1}{c}{{\specialcell[b]{Height}}} & \multicolumn{1}{c}{{\specialcell[b]{BMI}}} & \multicolumn{1}{c}{{\specialcell[b]{Fluid \\ intel.}}} & \multicolumn{1}{c}{{\specialcell[b]{Educ. \\ attain.}}}\\
\midrule
\addlinespace[0.75em]
\multicolumn{5}{l}{\textbf{Panel A -- Full sample}}\\
\midrule \hspace{1em}Breastfed & 0.697^{ *** } & -0.163^{ *** } & 0.116^{ *** } & 0.340^{ *** }\\
\hspace{1em} & (0.025) & (0.026) & (0.006) & (0.010)\\
\hspace{1em}Observations & \multicolumn{1}{D{,}{,}{-3}}{308,065} & \multicolumn{1}{D{,}{,}{-3}}{307,773} & \multicolumn{1}{D{,}{,}{-3}}{125,049} & \multicolumn{1}{D{,}{,}{-3}}{305,252}\\
\hspace{1em}Mean dep. var. & \multicolumn{1}{d}{168.577} & \multicolumn{1}{d}{27.367} & \multicolumn{1}{d}{0} & \multicolumn{1}{d}{13.163}\\
\hspace{1em}$R^2$ & \multicolumn{1}{d}{0.541} & \multicolumn{1}{d}{0.025} & \multicolumn{1}{d}{0.064} & \multicolumn{1}{d}{0.109}\\
\hspace{1em}District FE & {X} & {X} & {X} &  {X}\\
\addlinespace[0.75em]
\multicolumn{5}{l}{\textbf{Panel B -- Sibling sample}}\\
\midrule \hspace{1em}Breastfed & 0.617^{ *** } & -0.157^{ * } & 0.141^{ *** } & 0.379^{ *** }\\
\hspace{1em} & (0.104) & (0.086) & (0.026) & (0.034)\\
\hspace{1em}Observations & \multicolumn{1}{D{,}{,}{-3}}{23,251} & \multicolumn{1}{D{,}{,}{-3}}{23,233} & \multicolumn{1}{D{,}{,}{-3}}{8,712} & \multicolumn{1}{D{,}{,}{-3}}{23,152}\\
\hspace{1em}Mean dep. var. & \multicolumn{1}{d}{167.965} & \multicolumn{1}{d}{27.193} & \multicolumn{1}{d}{-0.006} & \multicolumn{1}{d}{13.065}\\
\hspace{1em}$R^2$ & \multicolumn{1}{d}{0.565} & \multicolumn{1}{d}{0.08} & \multicolumn{1}{d}{0.164} & \multicolumn{1}{d}{0.159}\\
\hspace{1em}District FE & {X} & {X} & {X} & {X}\\
\addlinespace[0.75em]
\multicolumn{5}{l}{\textbf{Panel C -- Sibling sample with family fixed effects}}\\
\midrule \hspace{1em}Breastfed & 0.315^{ ** } & -0.029^{  } & 0.120^{ ** } & 0.063^{  }\\
\hspace{1em} & (0.139) & (0.123) & (0.053) & (0.055)\\
\hspace{1em}Observations & \multicolumn{1}{D{,}{,}{-3}}{23,421} & \multicolumn{1}{D{,}{,}{-3}}{23,389} & \multicolumn{1}{D{,}{,}{-3}}{5,977} & \multicolumn{1}{D{,}{,}{-3}}{23,228}\\
\hspace{1em}Mean dep. var. & \multicolumn{1}{d}{167.979} & \multicolumn{1}{d}{27.182} & \multicolumn{1}{d}{-0.032} & \multicolumn{1}{d}{13.077}\\
\hspace{1em}$R^2$ & \multicolumn{1}{d}{0.889} & \multicolumn{1}{d}{0.652} & \multicolumn{1}{d}{0.705} & \multicolumn{1}{d}{0.717}\\
\hspace{1em}Family FE & {X} & {X} & {X} & {X}\\
\bottomrule
\end{tabular}
\begin{tablenotes}
\item Columns: (1) height in centimeters, (2) body mass index, (3) standardised fluid intelligence score, (4) years of education. Regresses outcomes onto an indicator for having been breastfed, controlling for sex and year-month of birth fixed effects, as well as either family or district fixed effects. (*): $p < 0.1$, (**): $p<0.05$, (***): $p<0.01$.
\end{tablenotes}
\end{threeparttable}
\end{table}

Panel B presents the results, but restricts the analysis to the sibling sample. Although this increases the standard errors substantially, the estimates remain similar (in size and significance) to those in the previous panel. The exception is BMI, which becomes significant only at the 10\% level, but remains of similar magnitude. 
For each of the outcomes, the estimated effects exceed the corresponding minimum detectable effect sizes at 80\% power, placing them within the range of effects our analysis is powered to detect.

Panel C additionally includes family fixed effects, exploiting variation in breastfeeding and the outcomes of interest \textit{within} families only. In general, this attenuates the estimates and further increases the standard errors. The estimates are therefore not significantly different from zero, except for height and fluid intelligence, which remain positive and significant. This suggests that family and environmental factors that are shared by siblings account for much of the association of having been breastfed with health and human capital outcomes. 

\subsection{Gene-environment interplay}
Before we investigate potential moderation of the breastfeeding effect by individuals' genetic ``predisposition'' (i.e., the $G \times E$ interaction effect), \autoref{tab:predictive-power-pgis} (\textit{SI}) confirms the predictive power of the outcome-specific PGIs in both the full sample and the sibling sample. The incremental $R^2$ values range from 6.5\% in the case of fluid intelligence (Panel B) to 14.8\% in the case of height (Panel A). In Panel C, where we estimate effect sizes \textit{within} families, all PGI coefficients (other than the education PGI) remain similar, but the incremental $R^2$ values are unsurprisingly smaller. The similar magnitudes suggest that the PGI predominantly captures direct (causal) genetic effects. The exception however, is educational attainment, for which the coefficient halves, consistent with the existing literature \citep{koellinger2018using, kong2018nature}. 

\autoref{tab:assoc_bf-genetics} (\textit{SI}) examines the role of gene-environment correlation ($rGE$), since its presence would complicate the interpretation of the $G \times E$ coefficient. For example, if breastfeeding is systematically more common among individuals with higher or lower PGIs, its coefficient partially captures a genetic component. Although we find evidence of $rGE$ in the full and sibling sample, it is eliminated once we include family fixed effects. This suggests that the within-family analysis allows us to identify true $G \times E$ interplay, rather than spurious $G \times G$ or $E \times E$ interactions.

Our $G \times E$ analysis focuses on the sibling sample. Using only within-family variation leads to a reduction in the impact of confounding and ensures that the PGI captures direct genetic effects. However, it also reduces the power of our analyses; \textit{SI} Appendix \ref{sec:power_analysis} discusses this in more detail and shows that we have relatively low power to identify gene-environment interaction effects of this magnitude. We therefore are more cautious in the discussion of our findings and interpret our estimates more as \textit{suggestive} evidence. 

\autoref{tab:v2_assoc_long-run-outcomes_gxe} presents the within-family estimates for $G$ (the PGI predictive of the outcome of interest), $E$ (breastfeeding) and $G\times E$. This shows that the PGI effects are significantly different from zero, and of the same order of magnitude as in \autoref{tab:predictive-power-pgis} (\textit{SI}). 
In line with the sibling fixed effects estimates presented in Panel C of \autoref{tab:v2_assoc_long-run-outcomes}, having been breastfed does not affect BMI or education, but it significantly increases height and fluid intelligence. Finally, the $G\times E$ estimates are largely indistinguishable from zero, suggesting that one's genetic ``predisposition'' does not moderate the impact of breastfeeding. 
In the case of height however, having been breastfed strengthens the genetic predisposition of being taller, albeit only at the 10\% level of statistical significance. In other words, this suggests that breastfeeding increases individuals' height by more among those with a high PGI for height. These results are robust to additionally accounting for interactions between the control variables and $G$ or $E$ (\autoref{tab:long-run_gxe_robustness}; \textit{SI}); following \citet{Keller2014}.

\FloatBarrier
\begin{table}[!h]
\centering\centering\centering
\caption{\label{tab:v2_assoc_long-run-outcomes_gxe}Associations between long-run outcomes, breastfeeding, and genetics.}
\centering
\begin{threeparttable}
\fontsize{10}{12}\selectfont
\setlength{\tabcolsep}{10pt}
\begin{tabular}[t]{ldddd}
\toprule
\multicolumn{1}{c}{\em{}} & \multicolumn{4}{c}{\em{Dependent variable:}} \\
\cmidrule(l{3pt}r{3pt}){2-5}
\multicolumn{1}{c}{} & \multicolumn{1}{c}{(1)} & \multicolumn{1}{c}{(2)} & \multicolumn{1}{c}{(3)} & \multicolumn{1}{c}{(4)} \\
\multicolumn{1}{c}{ } & \multicolumn{1}{c}{{\specialcell[b]{Height}}} & \multicolumn{1}{c}{{\specialcell[b]{BMI}}} & \multicolumn{1}{c}{{\specialcell[b]{Fluid \\ intel.}}} & \multicolumn{1}{c}{{\specialcell[b]{Educ. \\ attain.}}}\\
\midrule
Breastfed & 0.267^{ ** } & -0.009^{  } & 0.121^{ ** } & 0.081^{  }\\
 & (0.122) & (0.118) & (0.052) & (0.055)\\
PGI & 3.197^{ *** } & 1.690^{ *** } & 0.264^{ *** } & 0.346^{ *** }\\
 & (0.093) & (0.095) & (0.041) & (0.042)\\
Breastfed x PGI & 0.166^{ * } & -0.030^{  } & -0.024^{  } & 0.010^{  }\\
 & (0.097) & (0.100) & (0.044) & (0.045)\\
\midrule
Observations & \multicolumn{1}{D{,}{,}{-3}}{22,569} & \multicolumn{1}{D{,}{,}{-3}}{22,537} & \multicolumn{1}{D{,}{,}{-3}}{5,741} & \multicolumn{1}{D{,}{,}{-3}}{22,388}\\
Mean dep. var. & \multicolumn{1}{d}{167.992} & \multicolumn{1}{d}{27.178} & \multicolumn{1}{d}{-0.042} & \multicolumn{1}{d}{13.057}\\
$R^2$ & \multicolumn{1}{d}{0.919} & \multicolumn{1}{d}{0.68} & \multicolumn{1}{d}{0.721} & \multicolumn{1}{d}{0.723}\\
Family FE & {X} & {X} & {X} & {X}\\
\bottomrule
\end{tabular}
\begin{tablenotes}
\item Columns: (1) height in centimeters, (2) body mass index, (3) standardised fluid intelligence score, (4) years of education. Regresses outcome onto breastfeeding indicator, standardised PGI for the outcome (except for weight where we instead use the PGI for BMI), and the interaction between standardised PGI and breastfeeding. Controls for sex, and includes year-month of birth fixed effects and family fixed effects. Standard errors clustered by family. (*): $p < 0.1$, (**): $p<0.05$, (***): $p<0.01$.
\end{tablenotes}
\end{threeparttable}
\end{table}

In summary, our findings suggest that the relationships between breastfeeding, BMI and education are driven by family and environment factors that are shared between siblings. In contrast, the relationship between breastfeeding and height and fluid intelligence is robust to using only within-family variation, and -- in case of height -- the effect is stronger for those with a genetic ``predisposition'' to be taller. We report a range of additional analyses in \textit{SI} \autoref{app:extra_tabs_figs}, exploring the results of different heterogeneity analyses, functional form specifications, and robustness checks, where we focus on the within-family analyses unless otherwise specified. Despite breastfeeding being more common among those born in high SES areas, we show that the long-term (within-family) effects of breastfeeding are similar for those born in high and low SES districts (\autoref{tab:v2_assoc_by-ses}; \textit{SI}). We also show that our estimates for BMI from \autoref{tab:v2_assoc_long-run-outcomes_gxe} are not sensitive to different thresholds commonly used in the BMI distribution (i.e., healthy weight, overweight, obese; \autoref{tab:v2_assoc_bmi-nonlinear}; \textit{SI}). To account for potential differential investments in children by birth order \citep{muslimova2025nature}, we rerun our analyses additionally controlling for a dummy variable indicating if the individual was firstborn (\autoref{tab:assoc_long-run_birth-order-controls}; \textit{SI}), and find similar effects. 
Finally, \autoref{birth_spacing_robustness} (\textit{SI}) indicates that the estimates are similar when distinguishing between shorter or longer birth spacing between siblings.

\section{Discussion}
This paper shows how maternal breastfeeding rates plummeted during the mid-20$^{th}$ century, dropping from over 80\% of babies being breastfed in the late 1930s to just 40\% in the late 1960s. We explore the relationship between maternal breastfeeding and child long-term outcomes within this historical context, examining the association between having been breastfed and a range of outcomes highlighted as important in the literature, including height, BMI, years of schooling and fluid intelligence.

We highlight the social gradient in breastfeeding, with individuals born in areas characterised by a higher socio-economic status being substantially more likely to have been breastfed. This discrepancy in breastfeeding rates, in turn, may drive the observed correlation between breastfeeding and offspring outcomes. Indeed, the raw data shows that individuals who were breastfed on average are taller, have higher education and higher fluid intelligence scores. The sample of siblings, however, allows us to account for some of this confounding. Our analyses include family fixed effects, comparing individuals who were breastfed with their full siblings who were not. This begs the question of why mothers breastfeed one but not the other sibling, something that is likely to reflect differences in child endowment at birth, maternal time constraints, and/or maternal health \citep[see e.g.,][]{wolpin1995,delbono2012,muslimova2025nature}. We show that our analysis is robust to including birth order as a covariate but unfortunately, our data do not allow us to explore this issue in more detail. Instead, we highlight it as an area of interest for further research. Our within-family specification attenuates the raw correlation, although the effects of breastfeeding on height and fluid intelligence remain. The effects on height are quantitatively small, however. More specifically, we find that individuals who were breastfed are 0.3cm taller on average than their siblings who were not breastfed, and have a 10\% of a standard deviation higher fluid intelligence score. 

The gene-environment interplay analysis shows that most of the returns to breastfeeding do not differ between those with a higher or lower genetic ``predisposition''. Although these analyses are characterised by relatively low statistical power, the $G\times E$ estimates are close to zero, suggesting that there are no strong gene-environment interaction effects. The exception is height, where the impact of breastfeeding on height is larger among those with a higher genetic predisposition to be tall. Furthermore, with the exception of the PGI for years of education, we find that the predictive power of the PGIs is similar in the between-family and within-family analysis, suggesting that it predominantly captures direct (causal) genetic effects. 

Our estimates are quantitatively and qualitatively meaningful. The lasting impacts on adult height and fluid intelligence, even after controlling for sibling fixed effects and genetic endowments, emphasize the importance of breastfeeding. They also provide additional evidence consistent with recommendations that encourage mothers to breastfeed when making feeding decisions for their infants and support universal access to breastfeeding-supportive maternity care \citep[e.g.][]{perez2023, patnode2025}. 

Three key points regarding the interpretation of our results should be taken into account. First, since we are interested in the \textit{long-term} impacts of breastfeeding, we have to rely on historical breastfeeding data, and we show that the likelihood of being breastfed reduced substantially during our $\sim$40-year observation window. We currently live in a different environment, characterised by rates of initiation of breastfeeding of over 70\% in England \citep{BF2024}, but also by e.g., large falls in the weeks following birth especially in more deprived areas \citep{OHID2024}, as well as higher average maternal employment rates, better hygiene, availability of high-quality infant formula, etc. It would be interesting to link some of these environmental changes to the downward trend in breastfeeding. However, most reflect gradual changes rather than distinct, discontinuous events. The change in environmental conditions also implies it may be difficult to extrapolate our findings to the UK (or other advanced economies) today. However, using historical data (and with that, historical contexts) is the only way to estimate long-term impacts, and these are crucial in understanding the potential late-life impacts of early life exposures. 

Second, we use the UK Biobank, a data source that is known to be unrepresentative, with its participants on average being healthier and wealthier than the rest of the UK population \citep{Fry2017}. Our analysis shows a social gradient in breastfeeding rates that was already apparent (albeit smaller) in the 1940s. Given the high intergenerational persistence in wealth, the mothers of UK Biobank participants are likely to have been healthier and wealthier than the mothers of non-participants. These two facts suggest that we may over-estimate rates of breastfeeding compared to the rates in the UK population. This in turn may bias our estimates, though the use of sibling differences will attenuate this concern.

Third, our measure of maternal breastfeeding is based on a 40+ year recall, rather than a direct observation immediately after birth, or obtained through hospital birth records. This may introduce measurement error due to children not correctly recalling whether they were breastfed. Assuming this is classical measurement error, it is likely to attenuate our estimates, underestimating the returns to breastfeeding. Another reason for expecting our estimates to be a lower bound is due to selection. Since UK Biobank participants were assessed in 2006–2010, we implicitly condition on survival until then. In the presence of frailty selection, where fragile individuals are more likely to die prior to assessment leaving stronger survivors in the sample, our  estimates are likely to be attenuated.

Nevertheless, the data allow us to study breastfeeding patterns during a time in which no systematic breastfeeding data are available. The further addition of genetic data is unique, in particular for a sample of siblings, providing a novel opportunity to shed light on the stark trends in breastfeeding during this period, highlight heterogeneities in breastfeeding rates, and examine the nature-nurture interplay in the returns to breastfeeding.

\section{Material and  Methods}
We start by using the full UK Biobank sample to examine the association between breastfeeding and the outcomes of interest. We follow the literature and focus on individuals' health and human capital outcomes. For health, we are interested in the impact on individuals' BMI and height. For human capital, we are interested in individuals fluid intelligence and their years of education. The baseline empirical specification can be written as:
\begin{equation}
y_{ijt} = \theta_j + \gamma_t + \alpha \text{BF}_{ijt} + \beta X_{ijt} + \epsilon_{ijt} \label{eq:spec1}
\end{equation}
for individual $i$ born in Local Government District (henceforth: district) $j$ in year-month $t$. Districts are defined using historical area boundaries, of which we observe 1,472. \autoref{eq:spec1} defines individuals' outcome $y_{ijt}$ as a linear function of breastfeeding status ($\text{BF}_{ijt}$, equals one if breastfed, zero otherwise) and controls $X_{ijt}$, which include a binary indicator for being male, as well as district fixed effects $\theta_j$, accounting for systematic spatial differences across districts, and year-month of birth fixed effects, $\gamma_t$. The latter account for the fact that older individuals are more likely to be breastfed, but are also more likely to have worse health and lower levels of education. Hence, we control for highly flexible, non-linear trends in the variables of interest by time of birth. The parameter of interest $\alpha$ then captures the association between long-run outcomes and having been breastfed as an infant. We estimate \autoref{eq:spec1} by OLS separately for each of our outcomes, clustering standard errors by district.

Next, we restrict our analysis to the sibling sample and exploit only within-family variation by including family fixed effects:
\begin{equation}
y_{ift} = \zeta_f + \gamma_t + \alpha \text{BF}_{ift} + \beta X_{ift} + \epsilon_{ift} \label{eq:spec2}
\end{equation}
for individual $i$ born in year-month $t$ in family $f$. The family fixed effects, $\zeta_f$, control for all factors that affect the outcome that are \emph{shared} by siblings in a family. The parameter $\alpha$, capturing the association between breastfeeding and long-run outcomes, is thus identified from variation in outcomes \emph{within} families where one or more children have been breastfed and at least one child has not, reducing concerns about confounding by shared family factors (such as socio-economic status or the parental genetics). We estimate \autoref{eq:spec2} by OLS for each of our outcomes, clustering standard errors by family.

Given that the family fixed effects control for parental genetic variation, the inclusion of the child's PGI allows us to estimate direct (causal) genetic effects. For our final specification we additionally include a polygenic index ($PGI_{ift}^y$) that is specific to the outcome of interest, as a measure for individuals' genetic ``predisposition'' to the outcome $y_{ift}$, as well as the interaction between the PGI and breastfeeding status: 
\begin{equation}
y_{ift} = \zeta_f + \gamma_t + \alpha \text{BF}_{ift} + \delta \text{PGI}_{ift}^y + \eta \text{BF}_{ift} \times \text{PGI}_{ift}^y + \beta X_{ift} + \epsilon_{ift}. \label{eq:spec3}
\end{equation}
The parameter $\alpha$ has the same interpretation as above and may be biased if there are unobserved confounders that vary within families and affect both outcomes and breastfeeding status. $\delta$ captures the impact of a one standard deviation increase in one's genetic ``predisposition'' for outcome $y$. Note that the inclusion of family fixed effects implies we do not have to include the genetic principal components. Finally, $\eta$ captures the interaction effect between the genetic effect and having been breastfed. Its interpretation depends on whether there is any evidence of gene-environment correlation; in the absence of $rGE$, the estimate captures a true gene-environment interaction effect. \autoref{app:extra_tabs_figs} (\textit{SI}) shows no evidence of $rGE$ once we control for family fixed effects.

\paragraph{Acknowledgements.} We gratefully acknowledge financial support from the European Research Council (Starting Grant Reference 851725), and the European Union (project \#101073237 -- ESSGN). We thank participants at the 2025 European Social Science Genetics Network (ESSGN) conference for helpful comments and suggestions. This research has been conducted using the UK Biobank Resource under Application Number 74002. This work also uses data provided through www.visionofbritain.org.uk which is copyright of the Great Britain Historical GIS Project and the University of Portsmouth. The views expressed in this publication are those of the authors and do not necessarily reflect those of the European Union, MSCA Horizon Europe, or ESSGN. Neither the European Union, nor the granting authority or ESSGN can be held responsible for them.

\paragraph{Data, Material, and Software Availability.} District IDs will be made available through the UK Biobank website, in line with its policy. The UK Biobank data are only accessible upon payment of a fee. Researchers can apply for data access directly at \url{https://ukbiobank.ac.uk}. Our code is available on GitHub at \url{https://github.com/normark/bfukb}.

\newpage 
\singlespacing
\printbibliography
\newpage 
\doublespacing

\clearpage
\section*{\bfseries Online Appendix}
\appendix
\counterwithin{figure}{section}
\counterwithin{table}{section}
\counterwithin{equation}{section}

\FloatBarrier
\section{Data and descriptive statistics}

\subsection{Data and sample selection}\label{sec:data}
We use the UK Biobank to investigate the relationships between being breastfed and one's later life health/human capital outcomes, and we explore whether and how these main effects are moderated by one's genetic predisposition. The data include information on demographics, health, cognition, and education, as well as participants’ genotypes and biological samples. We perform the following sample selection. First, starting from \pvarConstructIndividualsNAll consented UK Biobank individuals, we restrict the sample to those with non-missing year-month and location of birth, leaving \pvarConstructIndividualsNNonmissingLocDate individuals. Second, we drop a small number of individuals born before 1937 and in 1971, removing \pvarAnalysisObsInSmallCohorts observations. Third, we keep only the \pvarAnalysisObsDroppingNoBF individuals that have reported their breastfeeding status at least once, and restrict to those with a birth location that can be located in a Local Government District (henceforth: district) in England or Wales, resulting in \pvarAnalysisObsDroppingNoDistrict observations. We refer to this final dataset as our `full sample'. For our within-family analyses, we construct a `sibling sample' by keeping only individuals from our full sample with at least one sibling also observed in the sample. After dropping \pvarAnalysisObsSiblingTwinsDropped twins, this results in \pvarAnalysisObsFamsSiblingSample families for a total of \pvarAnalysisObsSiblingSample individuals. Depending on the missingness of our outcome of interest, the sample size further reduces to between 125,090--308,073 observations for the full sample, and 8,954--23,465 for the sibling sample.

We define breastfeeding status by field \texttt{1677} of the UK Biobank, containing individuals' answers (``Yes'', ``No'', ``Don't know'', ``Prefer not to answer'') to the question ``Were you breastfed when you were a baby?''. We code ``Yes'' as unity, ``No'' as zero, and other answers as missing. If individuals were asked at multiple assessments, we use the earliest reported non-missing answer. Sex is defined by field \texttt{31} and year-month of birth by fields \texttt{34} and \texttt{52}. We use individuals' eastings and northings of birth to identify the district in which individuals were born, using the shapefiles from \url{www.VisionofBritain.com} and merge these into the UK Biobank. We observe 1,472 districts. For our heterogeneity analyses, we classify the socioeconomic status (SES) of individuals' birth districts by merging in district-level occupational data from the 1951 UK Census. This allows us to calculate the share of `low SES’ residents, defined as the share of the population that are not in professional, managerial, or technical occupations, and classify districts with a share above the median as `low SES'.

We consider four long-run outcomes. We start by exploring the associations with Body Mass Index (BMI) and height (in centimeters), all measured in adulthood at the time of individuals' UK Biobank assessment. When multiple assessments are available for the same individual, we use the earliest measurement, and we only consider individuals for whom we observe all three outcomes. Next, we examine educational attainment defined by mapping individuals’ qualifications to years of education according to the crosswalk in \autoref{tab:years_educ_definition}. Finally, we explore the associations with fluid intelligence, a problem solving score measured by a test taken during individuals' UK Biobank assessments. It measures logic and reasoning abilities independent of attained level of education. We only use the scores from the offline tests, and as the score does not have an interpretable unit, we standardise it to have zero mean and unit variance in our analysis sample. 

\begin{table}[h]
\caption{\label{tab:years_educ_definition}Mapping between qualifications and years of education.}
\centering
\begin{threeparttable}
\begin{tabular}{lc}
\toprule
Qualifications & Years of education \\ \midrule
College or university degree & 16 \\
A/AS levels $+$ NVQ/HND/HNC & 14 \\
A/AS levels $+$ Other professional qualifications & 15 \\
NVQ/HND/HNC & 13 \\
Other professional qualifications & 12 \\
A/AS levels & 13 \\
CSEs, GCSEs, or O levels & 11 \\
No qualifications & 10 \\ \bottomrule
\end{tabular}
\begin{tablenotes}
\item Columns: (1) the qualifications recorded in the UK Biobank, (2) the assigned years of education. A plus indicates that the individual must hold both of the specified qualifications simultaneously.
\end{tablenotes}
\end{threeparttable}
\end{table}

We use the polygenic indices from the PGI repository \citep{becker2021resource} that have been constructed for the subsample of siblings in the UK Biobank. This ensures that the GWAS discovery samples used for the PGIs do not overlap with the `sibling sample' we use for estimating genetic and gene-environment interaction effects. We choose the PGIs that are specific to each of our long-run outcomes. 
We standardise them to have zero mean and unit variance in our analysis samples.

\subsection{Descriptive statistics}\label{sec:descr}
\begin{figure}[!h]\caption{\label{fig:v2_long-run-by-bf_raw}Long-run outcomes by breastfeeding status}\centering\includegraphics[width=0.8\textwidth]{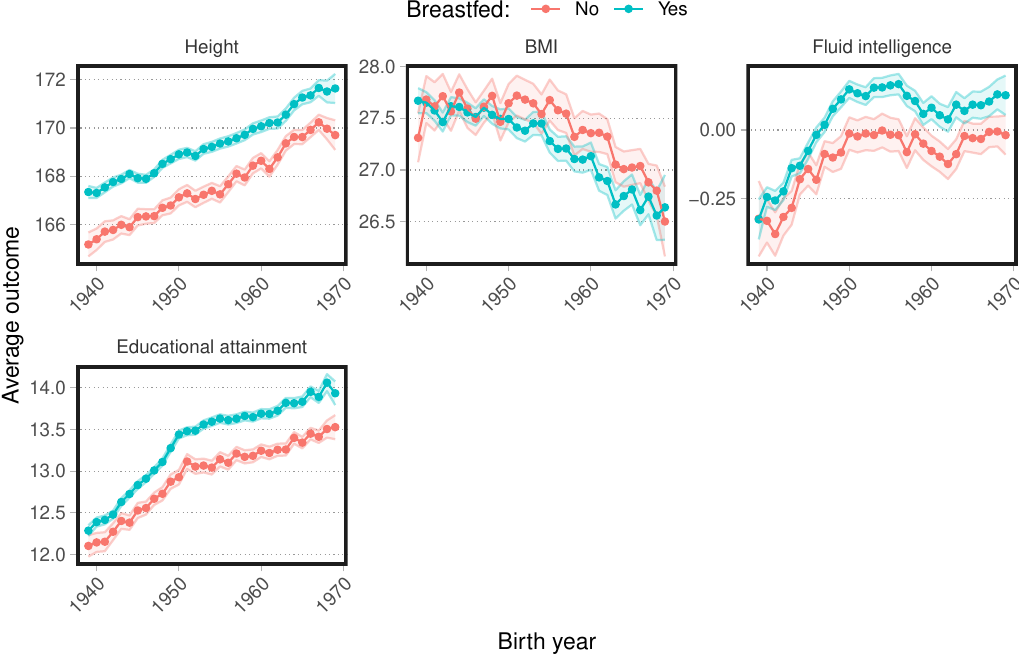}\caption*{\emph{Plots average long-run outcomes of UK Biobank participants by birth year and their breastfeeding status.}}\end{figure}

\begin{figure}[!h]\caption{\label{fig:bf-by-district}Breastfeeding by district}\centering\includegraphics[width=0.8\textwidth]{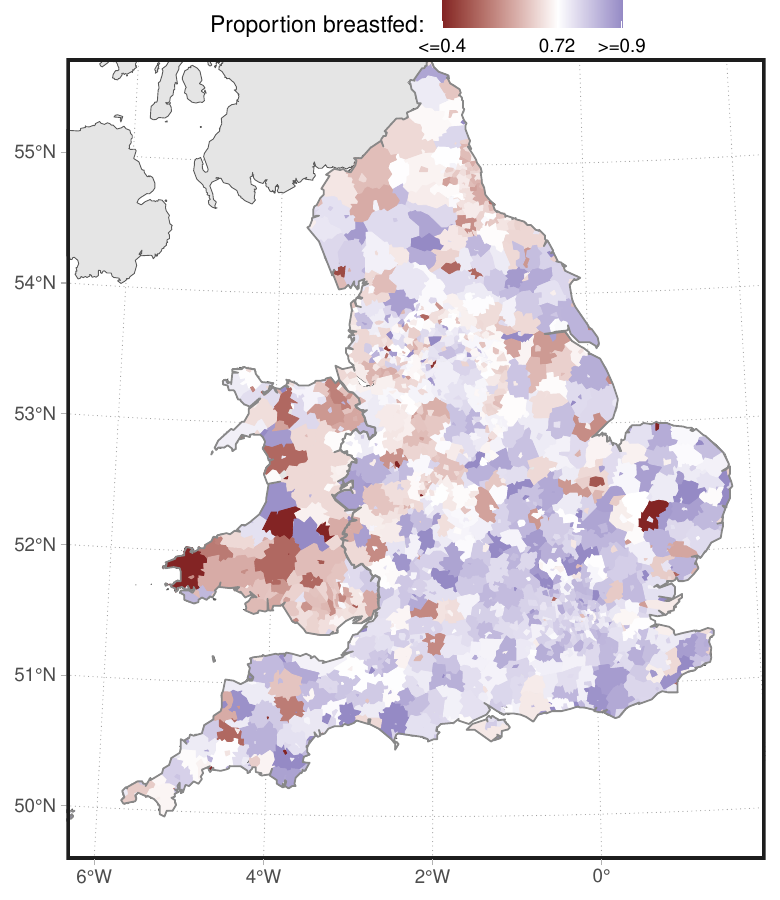}\caption*{\emph{Plots proportion of UK Biobank participants in England and Wales that were breastfed by district. Colored by deviation from the average taken across all districts in England and Wales (red is below average, blue is above).}}\end{figure}

\begin{figure}[!h]\caption{\label{fig:bf-by-ses}Breastfeeding by cohort and birth district SES}\centering\includegraphics[width=0.650\textwidth]{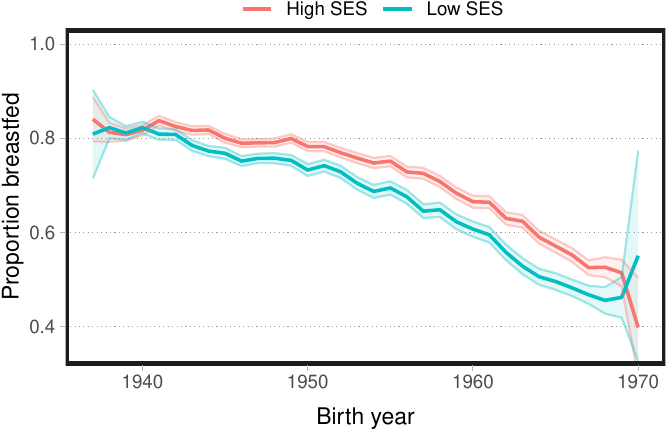}\caption*{\emph{Plots proportion of UK Biobank participants that were breastfed by the SES of their birth district and year-month of birth. We define low SES districts as those with an above median share of low SES residents by the 1951 UK Census, and high SES those with a below median share.}}\end{figure}

\autoref{fig:v2_long-run-by-bf_raw} shows that individuals who were breastfed are substantially taller, but with lower BMIs, and have higher intelligence and years of education; the latter two increasing with younger cohorts. We also find substantial regional variation in breastfeeding (\autoref{fig:bf-by-district}), highlighting wide variation in the probability of being breastfed across districts. Distinguishing between districts of relatively high and low socioeconomic status (SES) (defined by those with a below or above median level of low SES in the 1951 UK Census), we find substantial differences in breastfeeding rates. \autoref{fig:bf-by-ses} shows that the likelihood of breastfeeding reduces over time in both high and low SES districts, though the drop is steeper for the latter, increasing the SES gradient over time. 

The difference in breastfeeding rates between individuals born in high and low SES districts may partially be reflected in the raw associations between breastfeeding and child outcomes. We further explore the social gradient in breastfeeding by showing the association between breastfeeding and a range of other individual characteristics, measured (or fixed) at birth. This sheds light on the extent to which breastfeeding captures other individual or family characteristics. Using the sibling sample of the UK Biobank, Panel A of \autoref{tab:assoc_observables} shows that individuals who were breastfed have higher birth weights than children who were not, and their mothers were less likely to smoke around the birth of their child. On average, they also have higher PGIs for education and fluid intelligence, but we find no difference in the PGI for height or BMI. This again highlights the importance of confounding in the relationship between child outcomes and breastfeeding status, suggesting that, in a simple linear regression, breastfeeding is likely to capture other socioeconomic components, which may drive variation in the outcomes. 

Panel B of \autoref{tab:assoc_observables} shows the same analysis, but includes family fixed effects (instead of district fixed effects). This renders most of the differences insignificant. The exception is birth weight, which shows that higher birth weight siblings are more likely to have been breastfed. In other words, breastfeeding seems to be largely uncorrelated to biological predispositions, parity, and early maternal behavior, but it is positively associated with birth weight, suggesting that parents reinforce endowment differences. There is no difference in maternal smoking or any of the PGIs between siblings who were and those who were not breastfed. This suggests that the family fixed effects account for a substantial part of the confounding that is captured by the breastfeeding indicator. Including family fixed effects obviously also leads to a substantial reduction in the residual variation, since the coefficient is only identified using variation in breastfeeding between siblings in the same family. Our analysis shows that a regression of breastfeeding on the family fixed effects indeed captures \pvarAnalysisRsqBFbetween of the variation in breastfeeding, leaving \pvarAnalysisRsqBFwithin of the variation once family fixed effects have been controlled for.

\begin{landscape}
    \begin{table}[!h]
\centering\centering\centering
\caption{\label{tab:assoc_observables}Associations between observables and breastfeeding, between and within families.}
\centering
\begin{threeparttable}
\fontsize{10}{12}\selectfont
\setlength{\tabcolsep}{10pt}
\begin{tabular}[t]{lddddddd}
\toprule
\multicolumn{1}{c}{\em{}} & \multicolumn{7}{c}{\em{Dependent variable:}} \\
\cmidrule(l{3pt}r{3pt}){2-8}
\multicolumn{1}{c}{} & \multicolumn{1}{c}{(1)} & \multicolumn{1}{c}{(2)} & \multicolumn{1}{c}{(3)} & \multicolumn{1}{c}{(4)} & \multicolumn{1}{c}{(5)} & \multicolumn{1}{c}{(6)} & \multicolumn{1}{c}{(7)} \\
\multicolumn{1}{c}{ } & \multicolumn{1}{c}{{\specialcell[b]{Birth weight}}} & \multicolumn{1}{c}{{\specialcell[b]{Mother smoked}}} & \multicolumn{1}{c}{{\specialcell[b]{Firstborn}}} & \multicolumn{1}{c}{{\specialcell[b]{PGI, EA}}} & \multicolumn{1}{c}{{\specialcell[b]{PGI, FI}}} & \multicolumn{1}{c}{{\specialcell[b]{PGI, Height}}} & \multicolumn{1}{c}{{\specialcell[b]{PGI, BMI}}}\\
\midrule
\addlinespace[0.75em]
\multicolumn{8}{l}{\textbf{Panel A -- Sibling sample}}\\
\midrule \hspace{1em}Breastfed & 0.156^{ *** } & -0.078^{ *** } & -0.007^{  } & 0.090^{ *** } & 0.071^{ *** } & 0.020^{  } & -0.016^{  }\\
\hspace{1em} & (0.012) & (0.008) & (0.010) & (0.018) & (0.018) & (0.020) & (0.019)\\
\hspace{1em}Observations & \multicolumn{1}{D{,}{,}{-3}}{15,778} & \multicolumn{1}{D{,}{,}{-3}}{21,195} & \multicolumn{1}{D{,}{,}{-3}}{11,622} & \multicolumn{1}{D{,}{,}{-3}}{22,572} & \multicolumn{1}{D{,}{,}{-3}}{22,572} & \multicolumn{1}{D{,}{,}{-3}}{22,572} & \multicolumn{1}{D{,}{,}{-3}}{22,572}\\
\hspace{1em}Mean dep. var. & \multicolumn{1}{d}{3.345} & \multicolumn{1}{d}{0.275} & \multicolumn{1}{d}{0.383} & \multicolumn{1}{d}{-0.002} & \multicolumn{1}{d}{-0.002} & \multicolumn{1}{d}{-0.002} & \multicolumn{1}{d}{0.001}\\
\hspace{1em}$R^2$ & \multicolumn{1}{d}{0.101} & \multicolumn{1}{d}{0.097} & \multicolumn{1}{d}{0.162} & \multicolumn{1}{d}{0.109} & \multicolumn{1}{d}{0.083} & \multicolumn{1}{d}{0.076} & \multicolumn{1}{d}{0.079}\\
\hspace{1em}District FE & {X} & {X} & {X} & {X} & {X} & {X} & {X}\\
\addlinespace[0.75em]
\multicolumn{8}{l}{\textbf{Panel B -- Sibling sample with family fixed effects}}\\
\midrule \hspace{1em}Breastfed & 0.105^{ *** } & -0.006^{  } & 0.010^{  } & -0.016^{  } & -0.007^{  } & 0.011^{  } & 0.011^{  }\\
\hspace{1em} & (0.022) & (0.008) & (0.021) & (0.021) & (0.022) & (0.020) & (0.021)\\
\hspace{1em}Observations & \multicolumn{1}{D{,}{,}{-3}}{12,016} & \multicolumn{1}{D{,}{,}{-3}}{19,899} & \multicolumn{1}{D{,}{,}{-3}}{10,308} & \multicolumn{1}{D{,}{,}{-3}}{22,632} & \multicolumn{1}{D{,}{,}{-3}}{22,632} & \multicolumn{1}{D{,}{,}{-3}}{22,632} & \multicolumn{1}{D{,}{,}{-3}}{22,632}\\
\hspace{1em}Mean dep. var. & \multicolumn{1}{d}{3.355} & \multicolumn{1}{d}{0.254} & \multicolumn{1}{d}{0.387} & \multicolumn{1}{d}{-0.001} & \multicolumn{1}{d}{0} & \multicolumn{1}{d}{0.002} & \multicolumn{1}{d}{0}\\
\hspace{1em}$R^2$ & \multicolumn{1}{d}{0.708} & \multicolumn{1}{d}{0.89} & \multicolumn{1}{d}{0.629} & \multicolumn{1}{d}{0.785} & \multicolumn{1}{d}{0.768} & \multicolumn{1}{d}{0.781} & \multicolumn{1}{d}{0.761}\\
\hspace{1em}Family FE & {X} & {X} & {X} & {X} & {X} & {X} & {X}\\
\bottomrule
\end{tabular}
\begin{tablenotes}
\item Columns: (1) birth weight (kg), (2) whether mother smoked, (3) whether first born in the family, (4-7) PGIs for educ. attain, fluid intelligence, height, and BMI, respectively. Regresses observables onto an indicator for having been breastfed, year-month of birth fixed effects, as well as either district (Panel~A) or family (Panel~B) fixed effects. Uses the sibling sample. (*): $p < 0.1$, (**): $p<0.05$, (***): $p<0.01$.
\end{tablenotes}
\end{threeparttable}
\end{table}

\end{landscape}

\subsection{Power}\label{sec:power_analysis}
We calculate power in our pooled and within-family analyses using the following procedure.

\paragraph{Pooled design.} The data generating process is
\begin{equation}\label{eq:DGP}
y_i = \beta_0 + \beta_E E_i + \beta_G G_i + \beta_{GxE} G_i \times E_i + \epsilon_i    
\end{equation}
where $\epsilon_i \sim N(0, \sigma_\epsilon^2)$, $E_i \sim \text{Bernoulli}(p_E)$, $G_i \sim N(0, \sigma_G^2 = 1)$. We assume $E_i$ and $G_i$ to be independent, consistent with what we find for the within-family design (see \autoref{tab:assoc_bf-genetics} (\textit{SI})). We calculate the power for two-sided tests on the main effect  $\beta_E$ and the interaction effect $\beta_{GxE}$. The significance level is $\alpha = 0.05$. We list the values of parameters shared across outcomes in \autoref{tab:power_hyperparams}, and the estimates of the residual variance $\sigma_\epsilon^2$ for each outcome in the pooled column of \autoref{tab:tab:power_residual_var}. Under these simplifying assumptions, we have the following analytical power expressions:
\begin{align*}
\delta_E &= |\beta_E| \sqrt{\frac{N p_E (1-p_E)}{\sigma_\epsilon^2}}, \\
\delta_{GxE} &= |\beta_{GxE}| \sqrt{\frac{N p_E (1-p_E) \sigma_G^2}{\sigma_\epsilon^2}},\\
\text{Power}(\beta_k) &= 2 - \Phi(z_{1-\alpha/2} - \delta_k) - \Phi(z_{1-\alpha/2}  + \delta_k), \quad k \in \{E, GxE\}
\end{align*}
where $|\cdot|$ denotes absolute value and $z_{1-\alpha/2}$ is the $1-\alpha/2$ quantile from the standard normal distribution. We have verified these formulas by simulating our DGP based on the same assumptions (not shown). 

\paragraph{Sibling design.} We assume that families have size 2, such that our sample consists of sibling pairs. We then take sibling differences using Equation \ref{eq:DGP} to get:
\begin{equation}
\Delta y_i = \beta_E\Delta  E_i + \beta_G\Delta  G_i + \beta_{GxE}\Delta GE_i + \Delta \epsilon_i
\end{equation}
where $\Delta$ denotes sibling differences, so $\Delta y_i = (y_{i1} - y_{i2})$, $\Delta E_i = (E_{i1} - E_{i2})$, $\Delta G_i = (G_{i1} - G_{i2})$, $\Delta GE_i = (G_{i1} E_{i1} - G_{i2} E_{i2})$, $\Delta \epsilon_i = (\epsilon_{i1} - \epsilon_{i2})$, and $\Delta\epsilon_i \sim N(0, \sigma_{\Delta\epsilon}^2 = 2\sigma_\epsilon^2)$. We again assume independence  between $E_i$ and $G_i$, and also that
$$(G_{i1}, G_{i2}) \sim N(\mathbf{0}, \Sigma_G), \quad \Sigma_G = \begin{pmatrix} 1 & \rho \\ \rho & 1 \end{pmatrix}.$$
We calculate the power for two-sided tests on the main effect  $\beta_E$ and the interaction effect $\beta_{GxE}$. The significance level is $\alpha = 0.05$. 
\autoref{tab:power_hyperparams} lists the values of parameters shared across outcomes, with \autoref{tab:tab:power_residual_var} showing the estimates of the residual variance $\sigma_\epsilon^2$ for each outcome in the family column. 

Under these simplifying assumptions, we derive the analytical power formulas:
\begin{align*}
\delta_E &= |\beta_E|\sqrt{\frac{qN_{fam}}{2\sigma_\epsilon^2}}, \\
\delta_{GxE} &= |\beta_{GxE}|\sqrt{\frac{(2\,p_E(1-p_E)\,(1-\rho) + q\,\rho) N_{fam}}{2\sigma_\epsilon^2}}, \\
%\gamma &= 2\,p_E(1-p_E)\,(1-\rho) + q\,\rho, \\
\text{Power}(\beta_k) &= 2 - \Phi(z_{1-\alpha/2} - \delta_k) - \Phi(z_{1-\alpha/2} + \delta_k), \quad k \in \{E, GxE\}.
\end{align*}
where $z_{1-\alpha/2}$ is the $1-\alpha/2$ quantile from the standard normal distribution. Similar to above, we have verified these by simulating our DGP based on the same assumptions (not shown).

\begin{table}
\caption{\label{tab:power_hyperparams}Parameters for power calculations}
\centering
\begin{threeparttable}
\begin{tabular}[t]{llcl}
\toprule
 & Description & \multicolumn{1}{c}{Value} & Source \\
\midrule
$p_E$ & Share of breastfed siblings & 0.789 & \citet{colen2014breast} \\
$q$ & Share of discordant sibling pairs & 0.203 & \citet{rothstein2013breastfeeding} \\
$\rho$ & Within-family genetic correlation & 0.500 & Assume full siblings\\
$N$ & Number of siblings & 23000 &  Approx. size of our UKB sibling sample \\
$N_{fam}$ & Number of sibling pairs & 11500 & Assume 2 siblings per family \\
\bottomrule
\end{tabular}
\begin{tablenotes}
\item Lists parameter values we use in the power calculations and their source.
\end{tablenotes}
\end{threeparttable}
\end{table}

\begin{table}

\caption{\label{tab:tab:power_residual_var}Residual variances}
\centering
\begin{threeparttable}
\begin{tabular}[t]{rdd}
\toprule
 & \multicolumn{1}{c}{$\sigma_\epsilon^2$, family} & \multicolumn{1}{c}{$\sigma_\epsilon^2$, pooled}\\
\midrule
Educ. attain. & 2.990 & 5.194\\
Height & 57.502 & 82.712\\
BMI & 15.662 & 22.081\\
\bottomrule
\end{tabular}
\begin{tablenotes}
\item Lists for each outcome the residual variance $\sigma_\epsilon^2$ for the family and pooled designs. All variances were estimated in our sibling analysis sample.
\end{tablenotes}
\end{threeparttable}
\end{table}

\begin{figure}[!h]\caption{\label{fig:power_e}Power curves -- E effect.}\centering\includegraphics[width=\textwidth]{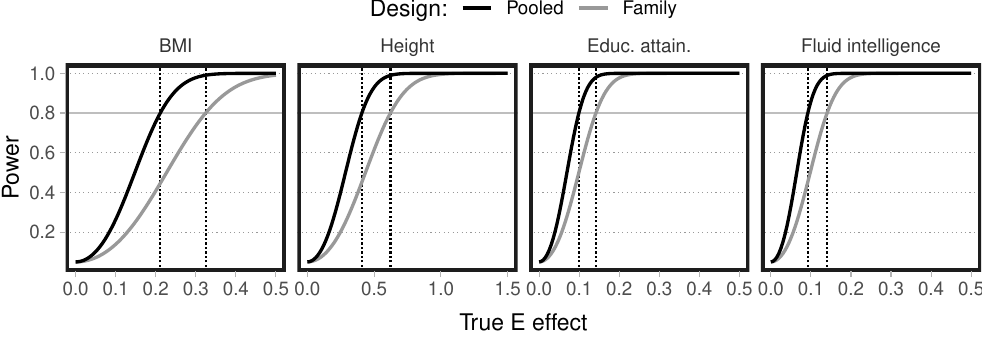}\caption*{\emph{Plots power as a function of true E effect size by design (pooled and family)}}\end{figure}

\autoref{fig:power_e} presents the power curves for the coefficient on breastfeeding $\beta_E$ using the sibling sample, whilst distinguishing between the pooled and within-family analysis. The dotted lines indicate the effect sizes we can reliably detect with 80\% power. This shows that, as expected, we have more power in the pooled analyses. The power curves suggest that, at 80\% power, we can reliably detect effect sizes as small as 0.21 (BMI), 0.41 (height), 0.10 (educational attainment), and 0.09 (fluid intelligence). In the within-family design, the corresponding detectable effect sizes are 0.33 (BMI), 0.62 (height), 0.14 (educational attainment and fluid intelligence). 

\begin{figure}[!h]\caption{\label{fig:power_gxe}Power curves -- GxE effect.}\centering\includegraphics[width=\textwidth]{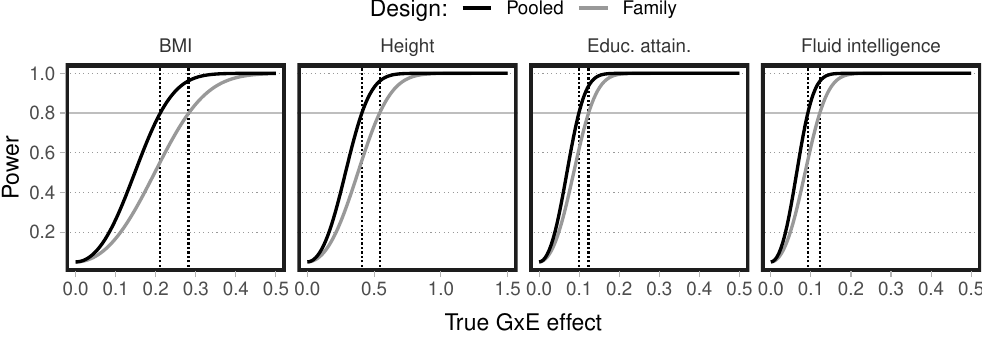}\caption*{\emph{Plots power as a function of true GxE effect size by design (pooled and family)}}\end{figure}

\autoref{fig:power_gxe} presents the power curves for the gene-environment interaction coefficient $\beta_{GxE}$, again based on the sibling sample and distinguishing between the pooled and within-family analysis. This shows that, at 80\% power, we can reliably detect effect sizes as small as 0.21 (BMI), 0.41 (height), 0.10 (educational attainment), and 0.09 (fluid intelligence). In the within-family design, the corresponding detectable effect sizes are 0.28 (BMI), 0.54 (height), 0.12 (educational attainment and fluid intelligence). 

\FloatBarrier
\clearpage 
\section{Additional Tables and Figures}
\label{app:extra_tabs_figs}

\begin{table}[!h]
\centering\centering\centering
\caption{\label{tab:predictive-power-pgis}Predictive power of polygenic indices.}
\centering
\begin{threeparttable}
\fontsize{10}{12}\selectfont
\setlength{\tabcolsep}{10pt}
\begin{tabular}[t]{ldddd}
\toprule
\multicolumn{1}{c}{\em{}} & \multicolumn{4}{c}{\em{Dependent variable:}} \\
\cmidrule(l{3pt}r{3pt}){2-5}
\multicolumn{1}{c}{} & \multicolumn{1}{c}{(1)} & \multicolumn{1}{c}{(2)} & \multicolumn{1}{c}{(3)} & \multicolumn{1}{c}{(4)} \\
\multicolumn{1}{c}{ } & \multicolumn{1}{c}{{\specialcell[b]{Height}}} & \multicolumn{1}{c}{{\specialcell[b]{BMI}}} & \multicolumn{1}{c}{{\specialcell[b]{Fluid \\ intel.}}} & \multicolumn{1}{c}{{\specialcell[b]{Educ. \\ attain.}}}\\
\midrule
\addlinespace[0.75em]
\multicolumn{5}{l}{\textbf{Panel A -- Full sample}}\\
\midrule \hspace{1em}PGI & 3.568^{ *** } & 1.662^{ *** } & 0.276^{ *** } & 0.636^{ *** }\\
\hspace{1em} & (0.012) & (0.009) & (0.003) & (0.006)\\
\hspace{1em}Observations & \multicolumn{1}{D{,}{,}{-3}}{288,574} & \multicolumn{1}{D{,}{,}{-3}}{288,318} & \multicolumn{1}{D{,}{,}{-3}}{116,266} & \multicolumn{1}{D{,}{,}{-3}}{287,245}\\
\hspace{1em}Mean dep. var. & \multicolumn{1}{d}{168.642} & \multicolumn{1}{d}{27.345} & \multicolumn{1}{d}{0} & \multicolumn{1}{d}{13.141}\\
\hspace{1em}$R^2$ & \multicolumn{1}{d}{0.691} & \multicolumn{1}{d}{0.146} & \multicolumn{1}{d}{0.137} & \multicolumn{1}{d}{0.179}\\
\hspace{1em}Incremental $R^2$ & 0.148 & 0.119 & 0.073 & 0.073\\
\hspace{1em}District FE & {X} & {X} & {X} &  {X}\\
\addlinespace[0.75em]
\multicolumn{5}{l}{\textbf{Panel B -- Sibling sample}}\\
\midrule \hspace{1em}PGI & 3.652^{ *** } & 1.678^{ *** } & 0.282^{ *** } & 0.664^{ *** }\\
\hspace{1em} & (0.041) & (0.035) & (0.010) & (0.018)\\
\hspace{1em}Observations & \multicolumn{1}{D{,}{,}{-3}}{22,538} & \multicolumn{1}{D{,}{,}{-3}}{22,520} & \multicolumn{1}{D{,}{,}{-3}}{8,413} & \multicolumn{1}{D{,}{,}{-3}}{22,445}\\
\hspace{1em}Mean dep. var. & \multicolumn{1}{d}{167.979} & \multicolumn{1}{d}{27.188} & \multicolumn{1}{d}{-0.006} & \multicolumn{1}{d}{13.046}\\
\hspace{1em}$R^2$ & \multicolumn{1}{d}{0.715} & \multicolumn{1}{d}{0.197} & \multicolumn{1}{d}{0.236} & \multicolumn{1}{d}{0.23}\\
\hspace{1em}Incremental $R^2$ & 0.149 & 0.116 & 0.068 & 0.076\\
\hspace{1em}District FE & {X} & {X} & {X} & {X}\\
\addlinespace[0.75em]
\multicolumn{5}{l}{\textbf{Panel C -- Sibling sample with family fixed effects}}\\
\midrule \hspace{1em}PGI & 3.320^{ *** } & 1.709^{ *** } & 0.266^{ *** } & 0.347^{ *** }\\
\hspace{1em} & (0.052) & (0.052) & (0.024) & (0.024)\\
\hspace{1em}Observations & \multicolumn{1}{D{,}{,}{-3}}{22,569} & \multicolumn{1}{D{,}{,}{-3}}{22,537} & \multicolumn{1}{D{,}{,}{-3}}{5,741} & \multicolumn{1}{D{,}{,}{-3}}{22,388}\\
\hspace{1em}Mean dep. var. & \multicolumn{1}{d}{167.992} & \multicolumn{1}{d}{27.178} & \multicolumn{1}{d}{-0.035} & \multicolumn{1}{d}{13.057}\\
\hspace{1em}$R^2$ & \multicolumn{1}{d}{0.919} & \multicolumn{1}{d}{0.679} & \multicolumn{1}{d}{0.721} & \multicolumn{1}{d}{0.723}\\
\hspace{1em}Incremental $R^2$ & 0.029 & 0.030 & 0.013 & 0.005\\
\hspace{1em}Family FE & {X} & {X} & {X} & {X}\\
\bottomrule
\end{tabular}
\begin{tablenotes}
\item Columns: (1) height in centimeters, (2) body mass index, (3) standardised fluid intelligence score, (4) years of education. Regresses outcome onto standardised PGI for the outcome (except for weight where we use the PGI for BMI). Controls for sex as well as year-month of birth fixed effects and family fixed effects. Standard errors clustered by family. (*): $p < 0.1$, (**): $p<0.05$, (***): $p<0.01$.
\end{tablenotes}
\end{threeparttable}
\end{table}

\begin{table}[!h]
\centering\centering\centering
\caption{\label{tab:assoc_bf-genetics}Associations between breastfeeding and genetics.}
\centering
\begin{threeparttable}
\fontsize{9}{11}\selectfont
\setlength{\tabcolsep}{5pt}
\begin{tabular}[t]{ldddd}
\toprule
\multicolumn{1}{c}{\em{}} & \multicolumn{4}{c}{\em{Dependent variable:}} \\
\cmidrule(l{3pt}r{3pt}){2-5}
\multicolumn{1}{c}{} & \multicolumn{1}{c}{(1)} & \multicolumn{1}{c}{(2)} & \multicolumn{1}{c}{(3)} & \multicolumn{1}{c}{(4)} \\
\multicolumn{1}{c}{ } & \multicolumn{1}{c}{{\specialcell[b]{Breastfed}}} & \multicolumn{1}{c}{{\specialcell[b]{Breastfed\hspace{0pt}}}} & \multicolumn{1}{c}{{\specialcell[b]{Breastfed\hspace{0pt}\hspace{0pt}}}} & \multicolumn{1}{c}{{\specialcell[b]{Breastfed\hspace{0pt}\hspace{0pt}\hspace{0pt}}}}\\
\midrule
\addlinespace[0.75em]
\multicolumn{5}{l}{\textbf{Panel A -- Full sample}}\\
\midrule \hspace{1em}PGI, fluid intelligence & 0.012^{ *** } & {} & {} &  {}\\
\hspace{1em} & (0.001) & {} & {} & {}\\
\hspace{1em}PGI, educ. attain. & {} & 0.014^{ *** } & {} & {}\\
\hspace{1em} & {} & (0.001) & {} & {}\\
\hspace{1em}PGI, height & {} & {} & 0.006^{ *** } & {}\\
\hspace{1em} & {} & {} & (0.001) & {}\\
\hspace{1em}PGI, BMI & {} & {} & {} & -0.007^{ *** }\\
\hspace{1em} & {} & {} & {} & (0.001)\\
\hspace{1em}Observations & \multicolumn{1}{D{,}{,}{-3}}{289,116} & \multicolumn{1}{D{,}{,}{-3}}{289,116} & \multicolumn{1}{D{,}{,}{-3}}{289,116} & \multicolumn{1}{D{,}{,}{-3}}{289,116}\\
\hspace{1em}Mean dep. var. & \multicolumn{1}{d}{0.72} & \multicolumn{1}{d}{0.72} & \multicolumn{1}{d}{0.72} & \multicolumn{1}{d}{0.72}\\
\hspace{1em}$R^2$ & \multicolumn{1}{d}{0.064} & \multicolumn{1}{d}{0.065} & \multicolumn{1}{d}{0.064} & \multicolumn{1}{d}{0.064}\\
\hspace{1em}District FE & {X} & {X} & {X} &  {X}\\
\addlinespace[0.75em]
\multicolumn{5}{l}{\textbf{Panel B -- Sibling sample}}\\
\midrule \hspace{1em}PGI, fluid intelligence & 0.012^{ *** } & {} & {} & {}\\
\hspace{1em} & (0.003) & {} & {} & {}\\
\hspace{1em}PGI, educ. attain. & {} & 0.015^{ *** } & {} & {}\\
\hspace{1em} & {} & (0.003) & {} & {}\\
\hspace{1em}PGI, height & {} & {} & 0.003^{  } & {}\\
\hspace{1em} & {} & {} & (0.003) & {}\\
\hspace{1em}PGI, BMI & {} & {} & {} & -0.003^{  }\\
\hspace{1em} & {} & {} & {} & (0.003)\\
\hspace{1em}Observations & \multicolumn{1}{D{,}{,}{-3}}{22,572} & \multicolumn{1}{D{,}{,}{-3}}{22,572} & \multicolumn{1}{D{,}{,}{-3}}{22,572} & \multicolumn{1}{D{,}{,}{-3}}{22,572}\\
\hspace{1em}Mean dep. var. & \multicolumn{1}{d}{0.776} & \multicolumn{1}{d}{0.776} & \multicolumn{1}{d}{0.776} & \multicolumn{1}{d}{0.776}\\
\hspace{1em}$R^2$ & \multicolumn{1}{d}{0.12} & \multicolumn{1}{d}{0.121} & \multicolumn{1}{d}{0.12} & \multicolumn{1}{d}{0.12}\\
\hspace{1em}District FE & {X} & {X} & {X} & {X}\\
\addlinespace[0.75em]
\multicolumn{5}{l}{\textbf{Panel C -- Sibling sample with family fixed effects}}\\
\midrule \hspace{1em}PGI, fluid intelligence & -0.001^{  } & {} & {} & {}\\
\hspace{1em} & (0.004) & {} & {} & {}\\
\hspace{1em}PGI, educ. attain. & {} & -0.003^{  } & {} & {}\\
\hspace{1em} & {} & (0.004) & {} & {}\\
\hspace{1em}PGI, height & {} & {} & 0.002^{  } & {}\\
\hspace{1em} & {} & {} & (0.004) & {}\\
\hspace{1em}PGI, BMI & {} & {} & {} & 0.002^{  }\\
\hspace{1em} & {} & {} & {} & (0.004)\\
\hspace{1em}Observations & \multicolumn{1}{D{,}{,}{-3}}{22,632} & \multicolumn{1}{D{,}{,}{-3}}{22,632} & \multicolumn{1}{D{,}{,}{-3}}{22,632} & \multicolumn{1}{D{,}{,}{-3}}{22,632}\\
\hspace{1em}Mean dep. var. & \multicolumn{1}{d}{0.777} & \multicolumn{1}{d}{0.777} & \multicolumn{1}{d}{0.777} & \multicolumn{1}{d}{0.777}\\
\hspace{1em}$R^2$ & \multicolumn{1}{d}{0.744} & \multicolumn{1}{d}{0.744} & \multicolumn{1}{d}{0.744} & \multicolumn{1}{d}{0.744}\\
\hspace{1em}Family FE & {X} & {X} & {X} & {X}\\
\bottomrule
\end{tabular}
\begin{tablenotes}
\item Columns show regressesions of breastfeeding status onto standardised PGIs (for fluid intelligence, educational attainment, height, and BMI). Includes year-month of birth fixed effects and district (Panels~A and B) or family (Panel~C) fixed effects. Standard errors clustered by district (Panels~A and B) or family (Panel~C). (*): $p < 0.1$, (**): $p<0.05$, (***): $p<0.01$.
\end{tablenotes}
\end{threeparttable}
\end{table}

\begin{table}[!h]
\centering\centering\centering
\caption{\label{tab:long-run_gxe_robustness}Associations between long-run outcome, breastfeeding, and genetics.}
\centering
\begin{threeparttable}
\fontsize{10}{12}\selectfont
\setlength{\tabcolsep}{10pt}
\begin{tabular}[t]{ldddd}
\toprule
\multicolumn{1}{c}{\em{}} & \multicolumn{4}{c}{\em{Dependent variable:}} \\
\cmidrule(l{3pt}r{3pt}){2-5}
\multicolumn{1}{c}{} & \multicolumn{1}{c}{(1)} & \multicolumn{1}{c}{(2)} & \multicolumn{1}{c}{(3)} & \multicolumn{1}{c}{(4)} \\
\multicolumn{1}{c}{ } & \multicolumn{1}{c}{{\specialcell[b]{Height}}} & \multicolumn{1}{c}{{\specialcell[b]{BMI}}} & \multicolumn{1}{c}{{\specialcell[b]{Fluid \\ intel.}}} & \multicolumn{1}{c}{{\specialcell[b]{Educ. \\ attain.}}}\\
\midrule
\addlinespace[0.75em]
\multicolumn{5}{l}{\textbf{Panel A -- G interacted with covariates, but not FE}}\\
\midrule \hspace{1em}Breastfed & 0.280^{ ** } & 0.012^{  } & 0.115^{ ** } & 0.093^{ * }\\
\hspace{1em} & (0.120) & (0.118) & (0.050) &  (0.055)\\
\hspace{1em}PGI & 3.175^{ *** } & 1.678^{ *** } & 0.269^{ *** } & 0.348^{ *** }\\
\hspace{1em} & (0.092) & (0.095) & (0.039) &  (0.042)\\
\hspace{1em}Breastfed x PGI & 0.180^{ * } & -0.029^{  } & -0.005^{  } & 0.011^{  }\\
\hspace{1em} & (0.096) & (0.099) & (0.042) &  (0.045)\\
\hspace{1em}Observations & \multicolumn{1}{D{,}{,}{-3}}{22,583} & \multicolumn{1}{D{,}{,}{-3}}{22,551} & \multicolumn{1}{D{,}{,}{-3}}{5,751} &  \multicolumn{1}{D{,}{,}{-3}}{22,402}\\
\hspace{1em}$R^2$ & \multicolumn{1}{d}{0.917} & \multicolumn{1}{d}{0.671} & \multicolumn{1}{d}{0.681} &  \multicolumn{1}{d}{0.715}\\
\addlinespace[0.75em]
\multicolumn{5}{l}{\textbf{Panel B -- G fully interacted}}\\
\midrule \hspace{1em}Breastfed & 0.290^{ ** } & 0.034^{  } & 0.114^{ ** } & 0.091^{ * }\\
\hspace{1em} & (0.120) & (0.118) & (0.050) & (0.055)\\
\hspace{1em}PGI & 3.179^{ *** } & 1.703^{ *** } & 0.261^{ *** } & 0.347^{ *** }\\
\hspace{1em} & (0.093) & (0.096) & (0.040) & (0.042)\\
\hspace{1em}Breastfed x PGI & 0.173^{ * } & -0.056^{  } & 0.002^{  } & 0.012^{  }\\
\hspace{1em} & (0.098) & (0.101) & (0.043) & (0.045)\\
\hspace{1em}Observations & \multicolumn{1}{D{,}{,}{-3}}{22,583} & \multicolumn{1}{D{,}{,}{-3}}{22,551} & \multicolumn{1}{D{,}{,}{-3}}{5,755} &  \multicolumn{1}{D{,}{,}{-3}}{22,402}\\
\hspace{1em}$R^2$ & \multicolumn{1}{d}{0.917} & \multicolumn{1}{d}{0.672} & \multicolumn{1}{d}{0.687} & \multicolumn{1}{d}{0.716}\\
\addlinespace[0.75em]
\multicolumn{5}{l}{\textbf{Panel C -- G fully interacted. E interacted with covariates, but not FE.}}\\
\midrule \hspace{1em}Breastfed & 0.281^{ ** } & 0.002^{  } & 0.116^{ ** } & 0.095^{ * }\\
\hspace{1em} & (0.121) & (0.118) & (0.051) & (0.055)\\
\hspace{1em}PGI & 3.176^{ *** } & 1.681^{ *** } & 0.269^{ *** } & 0.349^{ *** }\\
\hspace{1em} & (0.092) & (0.095) & (0.039) & (0.042)\\
\hspace{1em}Breastfed x PGI & 0.180^{ * } & -0.032^{  } & -0.005^{  } & 0.011^{  }\\
\hspace{1em} & (0.096) & (0.099) & (0.042) & (0.045)\\
\hspace{1em}Observations & \multicolumn{1}{D{,}{,}{-3}}{22,583} & \multicolumn{1}{D{,}{,}{-3}}{22,551} & \multicolumn{1}{D{,}{,}{-3}}{5,751} & \multicolumn{1}{D{,}{,}{-3}}{22,402}\\
\hspace{1em}$R^2$ & \multicolumn{1}{d}{0.917} & \multicolumn{1}{d}{0.671} & \multicolumn{1}{d}{0.681} & \multicolumn{1}{d}{0.715}\\
\addlinespace[0.75em]
\multicolumn{5}{l}{\textbf{Panel D -- G fully interacted. E fully interacted}}\\
\midrule \hspace{1em}Breastfed & 0.326^{ *** } & 0.013^{  } & 0.077^{  } & 0.089^{  }\\
\hspace{1em} & (0.121) & (0.120) & (0.051) & (0.057)\\
\hspace{1em}PGI & 3.190^{ *** } & 1.698^{ *** } & 0.272^{ *** } & 0.350^{ *** }\\
\hspace{1em} & (0.094) & (0.096) & (0.040) & (0.042)\\
\hspace{1em}Breastfed x PGI & 0.161^{  } & -0.051^{  } & -0.004^{  } & 0.010^{  }\\
\hspace{1em} & (0.098) & (0.101) & (0.044) & (0.046)\\
\hspace{1em}Observations & \multicolumn{1}{D{,}{,}{-3}}{22,583} & \multicolumn{1}{D{,}{,}{-3}}{22,551} & \multicolumn{1}{D{,}{,}{-3}}{5,755} & \multicolumn{1}{D{,}{,}{-3}}{22,402}\\
\hspace{1em}$R^2$ & \multicolumn{1}{d}{0.918} & \multicolumn{1}{d}{0.674} & \multicolumn{1}{d}{0.693} & \multicolumn{1}{d}{0.717}\\
\midrule
\hspace{1em}Mean dep. var. & \multicolumn{1}{d}{167.994} & \multicolumn{1}{d}{27.177} & \multicolumn{1}{d}{-0.041} & \multicolumn{1}{d}{13.058}\\
\hspace{1em}Family FE & {X} & {X} & {X} & {X}\\
\bottomrule
\end{tabular}
\begin{tablenotes}
\item Sibling sample with family fixed effects. We never include interactions with the family fixed effects. Columns: (1) height in centimeters, (2) body mass index, (3) standardised fluid intelligence score, (4) years of education. Regresses outcome onto breastfeeding indicator, standardised PGI for the outcome (except for weight where we instead use the PGI for BMI), controls (sex), and interactions. Includes year-month of birth fixed effects and family fixed effects. Standard errors clustered by family. (*): $p < 0.1$, (**): $p<0.05$, (***): $p<0.01$.
\end{tablenotes}
\end{threeparttable}
\end{table}

\begin{table}[!h]
\centering\centering\centering
\caption{\label{tab:v2_assoc_by-ses}Associations between long-run outcomes and breastfeeding.}
\centering
\begin{threeparttable}
\fontsize{10}{12}\selectfont
\setlength{\tabcolsep}{10pt}
\begin{tabular}[t]{lddddd}
\toprule
\multicolumn{1}{c}{\em{}} & \multicolumn{5}{c}{\em{Dependent variable:}} \\
\cmidrule(l{3pt}r{3pt}){2-6}
\multicolumn{1}{c}{} & \multicolumn{1}{c}{(1)} & \multicolumn{1}{c}{(2)} & \multicolumn{1}{c}{(3)} & \multicolumn{1}{c}{(4)} & \multicolumn{1}{c}{(5)} \\
\multicolumn{1}{c}{ } & \multicolumn{1}{c}{{\specialcell[b]{Height}}} & \multicolumn{1}{c}{{\specialcell[b]{Weight}}} & \multicolumn{1}{c}{{\specialcell[b]{BMI}}} & \multicolumn{1}{c}{{\specialcell[b]{Fluid \\ intel.}}} & \multicolumn{1}{c}{{\specialcell[b]{Educ. \\ attain.}}}\\
\midrule
\addlinespace[0.75em]
\multicolumn{6}{l}{\textbf{Panel A -- High SES}}\\
\midrule \hspace{1em}Breastfed & 0.418^{ ** } & 0.711^{  } & 0.125^{  } & 0.159^{ ** } & 0.089^{  }\\
\hspace{1em} & (0.200) & (0.525) & (0.178) & (0.074) & (0.080)\\
\hspace{1em}Observations & \multicolumn{1}{D{,}{,}{-3}}{11,417} & \multicolumn{1}{D{,}{,}{-3}}{11,407} & \multicolumn{1}{D{,}{,}{-3}}{11,403} & \multicolumn{1}{D{,}{,}{-3}}{2,866} & \multicolumn{1}{D{,}{,}{-3}}{11,334}\\
\hspace{1em}Mean dep. var. & \multicolumn{1}{d}{168.162} & \multicolumn{1}{d}{76.559} & \multicolumn{1}{d}{26.994} & \multicolumn{1}{d}{-0.054} & \multicolumn{1}{d}{13.19}\\
\hspace{1em}$R^2$ & \multicolumn{1}{d}{0.89} & \multicolumn{1}{d}{0.742} & \multicolumn{1}{d}{0.668} & \multicolumn{1}{d}{0.768} & \multicolumn{1}{d}{0.732}\\
\hspace{1em}Family FE & {X} & {X} & {X} & {X} &  {X}\\
\addlinespace[0.75em]
\multicolumn{6}{l}{\textbf{Panel B -- Low SES}}\\
\midrule \hspace{1em}Breastfed & 0.305^{  } & -0.073^{  } & -0.119^{  } & 0.250^{ *** } & 0.098^{  }\\
\hspace{1em} & (0.225) & (0.569) & (0.194) & (0.091) & (0.089)\\
\hspace{1em}Observations & \multicolumn{1}{D{,}{,}{-3}}{8,821} & \multicolumn{1}{D{,}{,}{-3}}{8,818} & \multicolumn{1}{D{,}{,}{-3}}{8,812} & \multicolumn{1}{D{,}{,}{-3}}{2,244} & \multicolumn{1}{D{,}{,}{-3}}{8,731}\\
\hspace{1em}Mean dep. var. & \multicolumn{1}{d}{167.451} & \multicolumn{1}{d}{77.326} & \multicolumn{1}{d}{27.491} & \multicolumn{1}{d}{-0.049} & \multicolumn{1}{d}{12.768}\\
\hspace{1em}$R^2$ & \multicolumn{1}{d}{0.894} & \multicolumn{1}{d}{0.747} & \multicolumn{1}{d}{0.669} & \multicolumn{1}{d}{0.741} & \multicolumn{1}{d}{0.714}\\
\hspace{1em}Family FE & {X} & {X} & {X} & {X} & {X}\\
\bottomrule
\end{tabular}
\begin{tablenotes}
\item Columns: (1) height in centimeters, (2) weight in kilograms, (3) body mass index, (4) standardised fluid intelligence score, (5) years of education. Regresses outcomes onto an indicator for having been breastfed, controlling for sex and year-month of birth fixed effects, as well as family fixed effects. (*): $p < 0.1$, (**): $p<0.05$, (***): $p<0.01$.
\end{tablenotes}
\end{threeparttable}
\end{table}

\begin{table}[!h]
\centering\centering\centering
\caption{\label{tab:v2_assoc_bmi-nonlinear}Associations between long-run BMI and breastfeeding -- non-linear.}
\centering
\begin{threeparttable}
\fontsize{10}{12}\selectfont
\setlength{\tabcolsep}{10pt}
\begin{tabular}[t]{lddd}
\toprule
\multicolumn{1}{c}{\em{}} & \multicolumn{3}{c}{\em{Dependent variable:}} \\
\cmidrule(l{3pt}r{3pt}){2-4}
\multicolumn{1}{c}{} & \multicolumn{1}{c}{(1)} & \multicolumn{1}{c}{(2)} & \multicolumn{1}{c}{(3)} \\
\multicolumn{1}{c}{ } & \multicolumn{1}{c}{{\specialcell[b]{Healthy weight \\ (18.5--25)}}} & \multicolumn{1}{c}{{\specialcell[b]{Overweight \\ (25--30)}}} & \multicolumn{1}{c}{{\specialcell[b]{Obese \\ ($\geq30$)}}}\\
\midrule
Breastfed & 0.000^{  } & 0.003^{  } & -0.003^{  }\\
 & (0.013) & (0.015) & (0.012)\\
\midrule
Observations & \multicolumn{1}{D{,}{,}{-3}}{23,389} & \multicolumn{1}{D{,}{,}{-3}}{23,389} & \multicolumn{1}{D{,}{,}{-3}}{23,389}\\
Mean dep. var. & \multicolumn{1}{d}{0.347} & \multicolumn{1}{d}{0.422} & \multicolumn{1}{d}{0.225}\\
$R^2$ & \multicolumn{1}{d}{0.595} & \multicolumn{1}{d}{0.536} & \multicolumn{1}{d}{0.594}\\
Family FE & {X} & {X} & {X}\\
\bottomrule
\end{tabular}
\begin{tablenotes}
\item Regresses outcomes onto breastfeeding indicator. Controls for sex, and includes year-month of birth fixed effects and family fixed effects. Standard errors clustered by family. (*): $p < 0.1$, (**): $p<0.05$, (***): $p<0.01$.
\end{tablenotes}
\end{threeparttable}
\end{table}

% BIRTH ORDER
\begin{table}[!h]
\centering\centering\centering
\caption{\label{tab:assoc_long-run_birth-order-controls}Associations between long-run outcomes and breastfeeding, controlling for being first born.}
\centering
\begin{threeparttable}
\fontsize{10}{12}\selectfont
\setlength{\tabcolsep}{10pt}
\begin{tabular}[t]{ldddd}
\toprule
\multicolumn{1}{c}{\em{}} & \multicolumn{4}{c}{\em{Dependent variable:}} \\
\cmidrule(l{3pt}r{3pt}){2-5}
\multicolumn{1}{c}{} & \multicolumn{1}{c}{(1)} & \multicolumn{1}{c}{(2)} & \multicolumn{1}{c}{(3)} & \multicolumn{1}{c}{(4)} \\
\multicolumn{1}{c}{ } & \multicolumn{1}{c}{{\specialcell[b]{Height}}} & \multicolumn{1}{c}{{\specialcell[b]{BMI}}} & \multicolumn{1}{c}{{\specialcell[b]{Fluid \\ intel.}}} & \multicolumn{1}{c}{{\specialcell[b]{Educ. \\ attain.}}}\\
\midrule
Breastfed & 0.292^{  } & -0.111^{  } & 0.119^{ ** } & -0.003^{  }\\
 & (0.219) & (0.196) & (0.058) & (0.086)\\
Firstborn & 0.020^{  } & 0.288^{ ** } & 0.032^{  } & 0.182^{ *** }\\
 & (0.141) & (0.131) & (0.038) & (0.057)\\
\midrule
Observations & \multicolumn{1}{D{,}{,}{-3}}{10,271} & \multicolumn{1}{D{,}{,}{-3}}{10,248} & \multicolumn{1}{D{,}{,}{-3}}{5,067} & \multicolumn{1}{D{,}{,}{-3}}{10,205}\\
Mean dep. var. & \multicolumn{1}{d}{168.263} & \multicolumn{1}{d}{27.005} & \multicolumn{1}{d}{-0.004} & \multicolumn{1}{d}{13.306}\\
$R^2$ & \multicolumn{1}{d}{0.893} & \multicolumn{1}{d}{0.675} & \multicolumn{1}{d}{0.712} & \multicolumn{1}{d}{0.724}\\
Family FE & {X} & {X} & {X} & {X}\\
\bottomrule
\end{tabular}
\begin{tablenotes}
\item Columns: (1) height in centimeters, (2) body mass index, (3) standardised fluid intelligence score, (4) years of education. Regresses outcomes onto an indicator for having been breastfed, controlling for sex, being first born, year-month of birth fixed effects, as well as family fixed effects. (*): $p < 0.1$, (**): $p<0.05$, (***): $p<0.01$.
\end{tablenotes}
\end{threeparttable}
\end{table}

\begin{figure}[!h]\caption{\label{birth_spacing_robustness}Long-run outcomes by breastfeeding status and birth spacing.}\centering\includegraphics[width=0.9\textwidth]{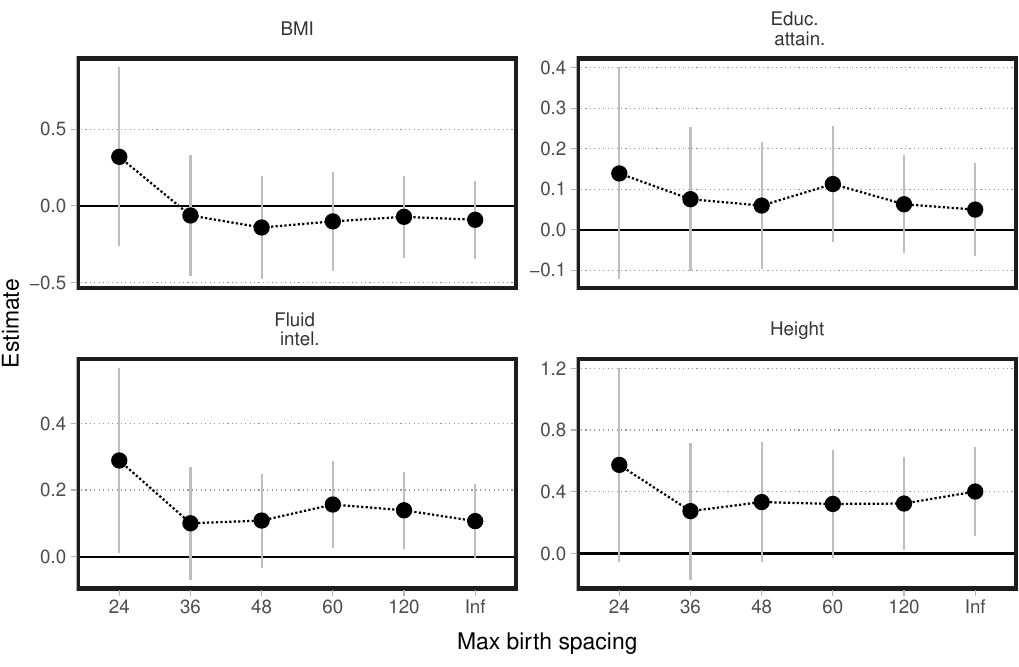}\caption*{\emph{Plots association between breastfeeding and long-run outcomes by threshold for maximum birth spacing. Vertical lines show 0.95 confidence intervals around the point estimates. Restricts the sample to families with two siblings in the UK Biobank.}}\end{figure}

\end{document}